\documentclass[]{raa}            
\usepackage{graphicx,times}
\usepackage{natbib}

\providecommand{\bm}[1]{\mbox{\boldmath$#1$\unboldmath}}

\begin{document}

   \title{Latest Cosmological Constraints on Cardassian expansion models including the updated Gamma-ray bursts
}

 \volnopage{ {\bf 2010} Vol.\ {\bf 10} No. {\bf XX}, 000--000}
   \setcounter{page}{1}

   \author{Nan Liang\inst{1}, Puxun Wu\inst{2}   \and Zong-Hong Zhu\inst{1}
        }

   \institute{Department of Astronomy, Beijing Normal   University, Beijing 100875, China \\
   {\it liangn@bnu.edu.cn; zhuzh@bnu.edu.cn}\\
        \and Center of Nonlinear Science and Department of Physics, Ningbo University,  No.818 Fenghua Road, Ningbo, Zhejiang, 315211
        China\\ {\it wpx0227@gmail.com}\\
\vs \no
   {\small Received [year] [month] [day]; accepted [year] [month] [day] }
}

\abstract{In this paper,  we constrain the Cardassian expansion
models  from the latest observations including the updated Gamma-ray
bursts (GRBs), which calibrated cosmology-independently from the
Union2 compilation of type Ia supernovae (SNe Ia). By combining the
GRB data to the joint observations with the Union2 SNe Ia set, along
with the Cosmic Microwave Background radiation observation from the
seven-year Wilkinson Microwave Anisotropy Probe result, the baryonic
acoustic oscillation observation from the spectroscopic Sloan
Digital Sky Survey  Data Release galaxy sample, we find significant
constraints on  model parameters of the original Cardassian model
$\Omega_{{\rm M0}}=0.282_{-0.014}^{+0.015}$, $n=
0.03_{-0.05}^{+0.05}$; and $n= -0.16_{-3.26}^{+0.25}$,
$\beta=0.76_{-0.58}^{+0.34}$ of the modified polytropic Cardassian
model, which are consistent with the $\Lambda$CDM model in
1-$\sigma$ confidence region. From the reconstruction of the
deceleration parameter $q(z)$ in Cardassian  models, we obtain the
transition redshift $z_{\rm T}=0.73\pm{0.04}$ for the original
Cardassian model, and $z_{\rm\rm T}=0.68\pm{0.04}$ for the modified
polytropic  Cardassian model.
  }

\authorrunning{N. Liang, P. Wu and Z.-H. Zhu}
\titlerunning{Constraints on the Cardassian expansion models including Gamma-ray bursts}
   \maketitle

\section{Introduction}
Recent years, the cosmological observations from type Ia supernovae
(SNe Ia; Riess et al.1998; Perlmutter et al. 1999), cosmic microwave
background radiation (CMB; Spergel et al. 2003), and large-scale
structures (LSS; Tegmark et al. 2004; Eisenstein et al. 2005) have
been used to explore cosmology extensively, which support that the
present expansion of our universe is accelerating. In order to
explain the accelerating expansion of the universe, many
cosmological models have been proposed. The first categories are
proposed by introducing an energy component called dark energy with
negative pressure in the universe, which dominates the universe to
drive the acceleration of expansion at recent times. Many candidates
of dark energy have been taken into account, such as the
cosmological constant with equation of state $w = -1$ (Carroll et
al. 1992), the scalar field models with dynamical equation of state,
e.g., quintessence (Ratra \& Peebles 1988; Caldwell et al. 1998),
phantom (Caldwell 2002), k-essence (Armendariz-Picon et al. 2001),
tachyon (Padmanabhan 2002, Sen 2005), quintom (Feng et al. 2005; Guo
et al. 2005;  Liang et al. 2009); as well as the Chaplygin gas
(Kamenshchik et al. 2001) and the generalized Chaplygin gas model
(GCG, Bento et al. 2002), the holographic dark energy (Cohen 1999;
Li 2004), the agegraphic dark energy (Cai 2008; Wei \&  Cai 2008a,
2008b), the Ricci dark energy (Gao et al. 2009) and so on. On the
other hand, many alternative models in which gravity is modified to
drive the universe acceleration have been proposed, e.g., the $f(R)$
theory in which the non-linear gravity Lagrangian ($L \sim R+f(R)$,
where $R$ is the scalar curvature) has been taken into account
(Capozziello \& Fang 2002); the braneworld models such as the
Dvali-Gabadadze-Porrati (DGP) model which consider that our
observable universe might be a surface or a brane embedded in a
higher dimensional bulk spacetime (Dvali et al. 2000); as well as
the Cardassian expansion model in which the Friedmann equation is
modified (Freese and Lewis 2002; Wang  et al. 2003).

In 2002, Freese and Lewis (Freese \&  Lewis 2002) proposed the
Cardassian expansion model as a possible explanation for the
acceleration by modifying the Friedmann equation without introducing
the dark energy.  The modified Friedmann equation  for the original
Cardassian model is
\begin{equation}
H^2=\frac{8\pi G}{3}(\rho+B\rho^n).
\end{equation}
The Cardassian term which is proportional to $\rho^n$ may show that
our observable universe as a $3+1$ dimensional brane is embedded in
extra dimensions.  The first term on the right side of the equation
dominates initially, so the equation becomes to the usual Friedmann
equation in the early universe. Then the two terms become equal at
redshift $ z=z_{\rm{card}}\sim O(1)$ (Freese \&  Lewis 2002), and
thereafter the Cardassian term begins to dominate the universe. $n$
is assumed to satisfy $n<2/3$ to give rise to a positive
acceleration of the universe. If $n=0$, the Cardassian term becomes
the cosmological constant. If $B=0$, the equation becomes the usual
FRW equation without the cosmological constant. Furthermore, the
modified polytropic Cardassian model can be obtained by introducing
an additional parameter $\beta$ into the original Cardassian model
(Wang  et al. 2003), and   the corresponding modified Friedmann
equation  is
\begin{equation}
H^2=\frac{8\pi G}{3}(\rho^\beta+C\rho^{n\beta})^{1/\beta}.
\end{equation}
When $\beta=1$, the modified polytropic Cardassian model reduces to
the original model.

Gamma-ray bursts (GRBs) are likely to occur in high-redshift range
beyond the SNe~Ia redshift limit. Up to now, the farthest GRB
detected is GRB 090423 at $z=8.2$ (Tanvir et al. 2009; Salvaterra et
al. 2009). Recently, several empirical GRB luminosity relations have
been proposed as distance indicators (Amati 2002; Norris et al.
2000; Fenimore \& Ramirez-Ruiz 2000; Reichart et al. 2001; Schaefer
2003a; Yonetoku et al. 2004; Ghirlanda et al. 2004a; Liang \& Zhang
2005; Firmani et al. 2006a; Yu et al. 2009). Therefore, GRBs could
be regarded as the standard candles to be a complementary
cosmological probe to the universe at high redshift (Schaefer 2003b;
Takahashi et al. 2003; Bloom et al.  2003,  Dai et al. 2004;
Ghirlanda et al. 2004b; Friedman \& Bloom 2005; Firmani et al. 2005,
2006b; Liang \& Zhang 2005;  Di Girolamo et al. 2005; Bertolami \&
Silva 2006; Ghirlanda et al. 2006; Schaefer 2007; Wright 2007; Wang
et al. 2007; Amati  et al. 2008; Basilakos \& Perivolaropoulos 2008;
Cuesta et al. 2008a, 2008b; Daly et al. 2008; Qi et al. 2008a,
2008b; Vitagliano et al. 2010). Due to the lack of the low-redshift
sample, these luminosity relations have been usually calibrated by
assuming a particular cosmological model (e.g., the $\Lambda$CDM
model with particular model parameters according to the concordance
cosmology). Therefore most of the calibration of GRBs are always
cosmology-dependent to lead the circularity problem in cosmological
research. Many of works treat the circularity problem with
statistical approaches such as simultaneous fitting of the
parameters in the calibration curves and the cosmology (Li et al.
2008; Wang 2008; Samushia, \& Ratra 2010;  Xu 2010;  Graziani 2010).
However, it is noted that an input cosmological model is still
required in doing the simultaneous fitting.

In our previous paper (Liang et al. 2008), we presented a new method
to calibrate GRB luminosity relations in a completely
cosmology-independent way. The luminosity distance of GRBs in the
redshift range of SNe Ia can be obtained by interpolating directly
from the SNe Ia Hubble diagram, and GRB data at high redshift can be
obtained by utilizing the calibrated relations (Liang et al. 2008).
Similar to the interpolation method, the luminosity distance of GRBs
could be obtained by other mathematical approach, such as the
empirical formula fitting (Kodama et al. 2008), the non-parametric
reconstruction method (Liang \& Zhang 2008), the local regression
(Cardone et al. 2009), and the cosmographic fitting (Gao et al.
2010; Capozziello \& Izzo 2010). Following the GRB calibration
method directly from SNe Ia, the derived cosmology-independent GRB
data at high redshift can be used to constrain cosmological models
by using the standard Hubble diagram method (Liang \& Zhang 2008;
Capozziello \& Izzo 2008; Izzo et al. 2009, Izzo \& Capozziello
2009; Wei \& Zhang 2009, Wei 2009;  Wang et al. 2009a, 2009b; Qi et
al. 2009; Wang \& Liang 2010; Liang et al. 2010; Wei 2010, Freitasa
et al. 2010; Liang et al. 2011; Liang \& Zhu 2011). Capozziello \&
Izzo (2008) first used the GRB relations calibrated with the
so-called Liang method to derive the cosmography parameters at high
redshift. Liang et al. (2010) combined the GRB data with the joint
data to constraint the cosmological parameters and reconstructed the
acceleration history of the universe.

Here we consider the Cardassian model viewed as purely
phenomenological modifications of the Friedmann equation to drive
the universe acceleration and  focus on the latest cosmological
constraints including GRBs. Up to now, the Cardassian model have
been constrained from many observational data, such as the angular
size of the compact radio sources (Zhu  \& Fujimoto 2002), SNe Ia
(Wang  et al. 2003; Zhu  \& Fujimoto 2003; Szydlowski  \& Czaja
2004; Godlowski et al. 2004; Frith  2004; Bento et al. 2005), the
x-ray gas mass fraction of clusters (Zhu  \& Fujimoto 2004; Zhu et
al. 2004), CMB (Sen \& Sen 2003; Savage et al. 2005), the large
scale structure (Multamaki et al. 2003; Amarzguioui et al. 2005; Fay
\& Amarzguioui 2006), the gravitational lensing (Alcaniz et al.
2005), the baryonic acoustic oscillation (BAO) (Wang et al. 2007),
the Hubble parameter versus redshift data (Yi \& Zhang 2007), as
well as the different combined data (Bento et al. 2006; Davis et al.
2007; Wang 2007;  Wang \& Wu 2009; Feng \& Li 2010). Also,
constraints from GRBs with the joint analysis on the Cardassian
model can be obtained in (Wang  et al. 2007; Cuesta et al. 2008a;
Wang  et al. 2009a; Wang \& Liang 2010). Very recently, the Union2
compilation of SNe Ia data set which consists of 557 SNe Ia has been
released (Amanullah et al. 2010), whereas the seven-year data of
Wilkinson Microwave Anisotropy Probe (WMAP7) has also been released
(Komatsu et al. 2010). In this paper, with the updated GRB data
calibrated directly from the Union2 set, we constrain the Cardassian
model and the modified polytropic Cardassian model from the latest
observations by combining the GRB
data to the joint observations with the Union2 set, 
along with the CMB observation  from the WMAP7 result, the BAO
observation from the spectroscopic Sloan Digital Sky Survey (SDSS)
Data Release  galaxy sample (Eisenstein et al. 2005). We also
reconstruct the deceleration parameter $q(z)$ in Cardassian
expansion models and obtain the transition redshift $z_{\rm T}$. We
find that tighter and more stringent constraints can be given out
with the combined data including GRBs in this work.

This paper is organized as follows. In section 2, we introduce the
analysis for the observation data. In section 3, we present
constraint results on Cardassian models from the joint observations
including GRBs, as well as SNe Ia, CMB, and BAO. Conclusions and
discussions are given in section 4.

\section{Observational Data Analysis}
In our previous paper (Liang et al. 2008; Liang et al. 2010), we
used the  192 SNe Ia compiled by Davis et al. (2007) and the 397 SNe
Ia set (Hicken et al. 2009) in  the interpolation procedure to
calibrate GRB luminosity relations from  the 69 GRBs compiled by
Schaefer (2007). A larger number of SNe Ia sample could bring more
accurate result in the interpolation procedure. Very recently, the
Union2 compilation (Amanullah et al. 2010) of 557 SNe Ia data set
has been released by the Supernova Cosmology Project Collaboration
(SCP). In this paper, we use the Union2 set to calibrate GRB
luminosity relations with the GRB sample at $z\le1.4$ by using the
linear interpolation method, and we update the distance moduli of
the GRBs at $z>1.4$ obtained by utilizing the new calibrated
relations. For more details for the calculation, see Liang et al.
(2008); Liang et al.(2010). We plot the Hubble diagram of Union2 SNe
Ia and the GRBs obtained using the interpolation methods in figure
1. The distance moduli of the 27 GRBs at $z\le1.4$ are obtained by
using the linear interpolation method directly from the Union2 SNe
set; the 42 GRB data at $z>1.4$ are obtained by utilizing the five
relations calibrated with the sample at $z\le1.4$.

 \begin{figure}
 \centering
\includegraphics[angle=0,width=0.6\textwidth]{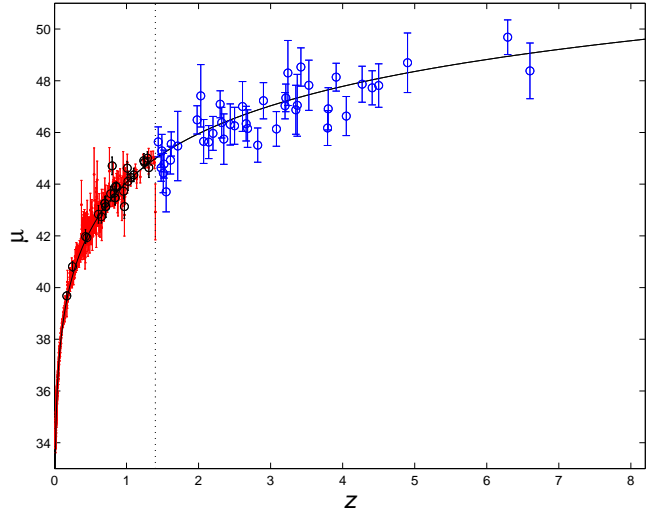}
\caption{Hubble Diagram of 557 SNe Ia (\textit{red dots}) and the 69
GRBs (\textit{circles}) obtained using the interpolation method. The
27 GRBs at $z\le1.4$ are obtained by linear interpolating from SNe
Ia data (\textit{black circles}), and the 42 GRBs at $z>1.4$
(\textit{blue circles}) are obtained with the five relations
calibrated with the sample at $z\le1.4$. The curve is the
theoretical distance modulus in the concordance model ($w= -1$,
$\Omega_{\rm M0}= 0.27$), and the vertical dotted line represents
$z=1.4$. } \label{fHD}
\end{figure}

The position of the first acoustic peak in the power spectrum of CMB
favors a spatially flat Universe, therefore we assume a flat
universe prior throughout this work. Constraints from  SNe Ia  and
GRB data can be obtained by fitting the distance moduli $\mu(z)$. A
distance modulus can be calculated as
\begin{eqnarray}\label{mu}
\mu=5\log \frac{d_L}{\rm{Mpc}} + 25=5\log_{10}D_L-\mu_0,
\end{eqnarray}
where $\mu_0=5\log_{10}h+42.38$, $h=H_0/(100{\rm km/s/Mpc})$, $H_0$
is the Hubble constant. For a flat universe, the luminosity distance
$D_L$ can be calculated by
\begin{eqnarray}\label{dLH}
D_L\equiv H_0d_L=(1+z)\int_0^z\frac{dz'}{E(z')},
\end{eqnarray}
where  $E(z)=H/H_0$, which determined by the choice of the specific
cosmological model.
The $\chi^2$ value of the observed distance moduli can be calculated
by
\begin{eqnarray}\label{chi2}
\chi^2_{\mu}=\sum_{i=1}^{N}\frac{[\mu_{\rm{obs}}(z_i)-\mu(z_i)]^2}
{\sigma_{\mu,i}^2},
\end{eqnarray}
where $\mu _{\rm obs}(z_i)$ are the observed distance modulus for
the SNe Ia and/or GRBs at redshift ~$z_i$~ with its error
$\sigma_{\mu_{\rm i}}$; $\mu(z_i)$ are the theoretical value of
distance modulus from cosmological models. Following the effective
approach (Nesseris \& Perivolaropoulos 2005), we marginalize the
nuisance parameter $\mu_0$ by minimizing
\begin{eqnarray}
{\hat\chi}^2_{\mu}=A- {B^2}/{C},
\end{eqnarray}
where
$A=\sum{[\mu_{\rm{obs}}(z_i)-5\log_{10}D_L]^2}/{\sigma_{\mu_{i}}^2}$,
$B=\sum{[\mu_{\rm{obs}}(z_i)-5\log_{10}D_L]}/{\sigma_{\mu_{i}}^2}$,
and $C=\sum{1}/{\sigma_{\mu_{i}}^2}$.

For the CMB observation, the shift parameters $R$ provide an
efficient summary of CMB data to constrain cosmological models. For
a flat universe, the shift parameter can be expressed as (Bond et
al. 1997)
\begin{equation}
R=\sqrt{\Omega_{\rm{M0}}}\int_0^{z_{\rm rec}}\frac{dz}{E(z)}
\end{equation}
here $z_{\rm rec}$ is the redshift of recombination. From the WMAP7
result, the shift parameter is constrained to be $R=1.725\pm0.018$
and $z_{\rm rec}=1091.3$ (Komatsu et al. 2010). The $\chi^2$ value
of the shift parameter can be calculated by
\begin{equation}
\chi^2_{\rm{CMB}}=\frac{(R-1.725)^2}{0.018^2}.
\end{equation}

For the BAO observation, we use the distance parameter $A$ which can
be expressed as for a flat universe (Eisenstein et al. 2005)
\begin{equation}
A=\frac{\sqrt{\Omega_{\rm{M0}}}}{E(z_{\rm
BAO})^{1/3}}[\frac{1}{z_{\rm BAO}}\int_0^{z_{\rm
BAO}}\frac{dz}{E(z)}]^{2/3}
\end{equation}
where $z_{\rm{BAO}}=0.35$. From the SDSS spectroscopic sample of
luminous red galaxy, the distance parameter is measured to be $A =
0.469(n_s/0.98)^{-0.35}\pm0.017$ (Eisenstein et al. 2005) , with the
scalar spectral index $n_s=0.963$ from the WMAP7 data (Komatsu et
al. 2010). The $\chi^2$ value of the distance parameter  can be
calculated by
\begin{equation}
\chi^2_{\rm{BAO}}=\frac{(A-0.4666)^2}{0.017^2}.
\end{equation}

\section{CONSTRAINTS FROM COMBINING GRBs, SNe Ia, CMB, and BAO }
In order to combine GRB data into the joint observational data
analysis to constrain cosmological models, we follow the simple way
to avoid any correlation between the SNe Ia data and the GRB data
(Liang et al. 2010).  The 40 SNe points used in the interpolating
procedure are excluded  from the Union2 SNe Ia sample used to the
joint constrains. The best fit values for model parameters can be
determined by minimizing
\begin{equation}
\chi^2=\hat\chi^2_{\mu\{\rm{517SNe+42GRBs}\}}+\chi^2_{\rm{CMB}}+\chi^2_{\rm{BAO}}\;.
\end{equation}

From the modified Friedmann equation of the original Cardassian
model, if only considering the matter term without considering the
radiation for simplification, using $\rho_{\rm
M}=\rho_{\rm{M0}}(1+z)^3=\Omega_{\rm{M0}}\rho_c(1+z)^3 $, where the
present  critical density of the universe $\rho_c=3H_0^2/8\pi G$, we
obtain
\begin{equation}
f_{\rm X}(z)\equiv\frac{\rho_{\rm{X}}}{\rho_{\rm{X0}}}=(1+z)^{3n},
\end{equation}
where subscript `x' refers to any component providing additional
term in the Friedmann equation. The corresponding $E(z)$ of the
Cardassian model is
\begin{eqnarray}\label{Ez}
E(z)=[\Omega_{\rm{M0}}(1+z)^3+(1-\Omega_{\rm{M0}})(1+z)^{3n}]^{1/2}.
\end{eqnarray}
For the modified polytropic Cardassian model, we obtain
\begin{equation}
f_X(z)=\frac{\Omega_{\rm{M0}}}{1-\Omega_{\rm{M0}}}(1+z)^3[(1+\frac{\Omega_{\rm{M0}}^{-\beta}-1}{(1+z)^{3(1-n)\beta}})^{1/\beta}-1]
\end{equation}
The corresponding $E(z)$ of the modified polytropic Cardassian model
is
\begin{equation}
E(z)=[\Omega_{\rm{M0}}(1+z)^3[1+\frac{\Omega_{\rm{M0}}^{-\beta}-1}{(1+z)^{3(1-n)\beta}}]^{1/\beta}]^{1/2}
\end{equation}

 \begin{table}[tbhp]
 \begin{center}{\scriptsize
 \begin{tabular}{|c|c|c|c|c|c|c|} \hline\hline
 & \multicolumn{3}{c|}{Original Cardassian Model}
& \multicolumn{3}{c|}{Modified polytropic Cardassian Model}  \\
 \cline{2-4} \cline{5-7}  &     SN+GRB+CMB+BAO                            &     SN+CMB+BAO               &     GRB+CMB+BAO                   &     SN+GRB+CMB+BAO                      &     SN+CMB+BAO                    &  GRB+CMB+BAO\\ \hline
 $\Omega_{\rm{M0}}$\ \ & \ \ $0.282_{-0.014}^{+0.015}$\ \  & \ \ $0.270_{-0.014}^{+0.014}$\ \ & \ \ $0.290_{-0.046}^{+0.045}$\ \         & \ \ $0.285_{-0.015}^{+0.014}$\ \        & \ \ $0.271_{-0.015}^{+0.015}$\ \  & \ \ $0.285_{-0.045}^{+0.045}$ \\
 $n$                   \ \ & \ \ $0.03_{-0.05}^{+0.05}$\ \  & \ \ $0.00_{-0.05}^{+0.05}$\ \   & \ \ $0.11_{-0.25}^{+0.18}$\ \             & \ \ $ -0.16_{-3.26}^{+0.25}$\ \           & \ \ $-0.22_{-3.27}^{+0.34}$\ \      & \ \ $-0.05_{-5.11}^{+0.59}$ \\
 $\beta$                   \ \ & \ \ $ \beta\equiv1$\ \     & \ \ $\beta\equiv1$\ \ & \ \ $\beta\equiv1$\ \                                   & \ \ $ 0.76_{-0.58}^{+0.34}$\ \         & \ \ $0.74_{-0.56}^{+1.15}$\ \      & \ \ $0.81_{-0.51}^{+3.80}$ \\
 $\chi_{\rm min}^2$    \ \ & \ \ $538.10$\ \                                       & \ \ $542.92$\ \                 & \ \ $34.76$\ \      & \ \ $537.46$\ \                       & \ \ $542.81$\ \                  & \ \ $34.76$  \\
 $\chi_{\rm min}^2/\rm{dof}$\ \ & \ \ $0.96$\ \                                & \ \ $0.98$\ \                    & \ \ $0.83$\ \      & \ \ $0.96$\ \                         & \ \ $0.98$\ \                    & \ \ $0.83$ \\
 \hline\hline
 \end{tabular} }
 \end{center}
 \caption{\label{tab2} The best-fit value of  parameters   $\Omega_{\rm{M0}}$,  $n$ and  $\beta$ for the original Cardassian model and the modified polytropic Cardassian model with $1\sigma$ uncertainties, as well as $\chi_{\rm min}^2$, $\chi_{\rm
min}^2/\rm{dof}$,  with SNe+GRBs+CMB+BAO, SNe+CMB+BAO, and
GRBs+CMB+BAO.}
 \end{table}

\begin{figure}
\begin{center}
\includegraphics[angle=0,scale=0.55]{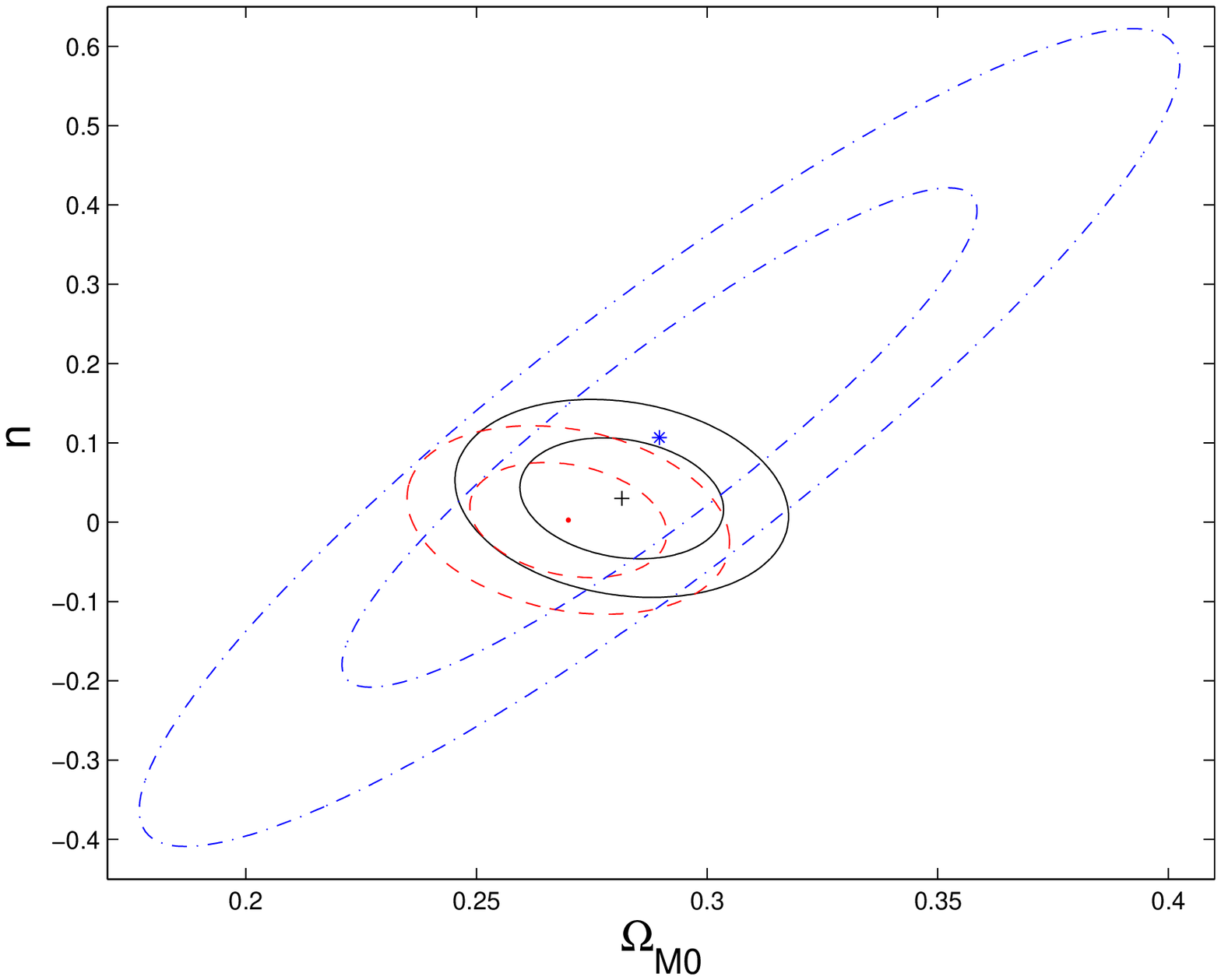}
\end{center}
\caption{The joint confidence regions in the $\Omega_{\rm{M0}}-n$
plane for the original  Cardassian model in a flat universe. The
contours correspond to 1-$\sigma$ and 2-$\sigma$ confidence regions.
The black solid lines, red dashed lines, the blue dash-dotted lines,
represent  the results of SNe+GRBs+CMB+BAO, SNe+CMB+BAO and
GRBs+CMB+BAO, respectively. The black plus, red point, and blue star
correspond the best-fit values of SNe+GRBs+CMB+BAO, SNe+CMB+BAO and
GRBs+CMB+BAO, respectively.}
\end{figure}

\begin{figure}
\begin{center}
\includegraphics[angle=0,scale=0.55]{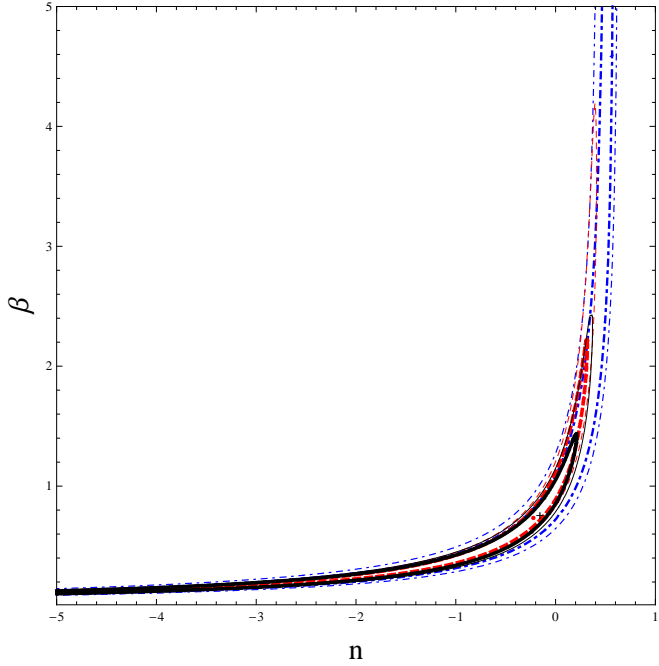}
\end{center}
\caption{The joint confidence regions in the $n-\beta$ plane for the
 modified polytropic Cardassian model in a flat universe.
 The contours correspond to 1-$\sigma$ and  2-$\sigma$ confidence regions. The black solid
lines, red dashed lines, the blue dash-dotted lines, represent  the
results of SNe+GRBs+CMB+BAO, SNe+CMB+BAO and GRBs+CMB+BAO,
respectively. The black plus, red point, and blue star correspond
the best-fit values of SNe+GRBs+CMB+BAO, SNe+CMB+BAO and
GRBs+CMB+BAO, respectively.}
\end{figure}

The joint confidence regions in $\Omega_{\rm{M0}}-n$ plane with the
combined observational data for the original Cardassian expansion
model are showed in figure 2. With SNe Ia + GRBs + CMB + BAO, the
best-fit values  at 1-$\sigma$ confidence level are
$\Omega_{\rm{M0}}=0.282_{-0.014}^{+0.015}$, $n=
0.03_{-0.05}^{+0.05}$. For comparison, fitting results from the
joint data without GRBs are also given in figure 2. The best-fit
values with SNe Ia + CMB + BAO are
$\Omega_{\rm{M0}}=0.270_{-0.014}^{+0.014}$, $n=
0.00_{-0.05}^{+0.05}$. While with GRBs + CMB + BAO without SNe Ia,
the best-fit values are $\Omega_{\rm{M0}}=0.290_{-0.046}^{+0.045}$,
$n= 0.11_{-0.25}^{+0.18}$. We present the best-fit value of
$\Omega_{\rm{M0}}$, $n$ with 1-$\sigma$ uncertainties, and
$\chi_{\rm min}^2$, $\chi_{\rm min}^2/\rm{dof}$ for the original
Cardassian model in Table 1.

For the modified polytropic Cardassian model, we find that the
best-fit values  at 1-$\sigma$ confidence level with SNe Ia + GRBs +
CMB + BAO are $\Omega_{\rm{M0}}=0.285_{-0.014}^{+0.015}$, $n=
-0.16_{-3.26}^{+0.25}$, $\beta=0.76_{-0.58}^{+0.36}$. Figure 3 shows
the joint confidence regions with the combined observational data
for the  modified polytropic Cardassian model in the $n-\beta$
plane, while fixing $\Omega_{\rm{M0}}$ with the best-fit values.
With SNe Ia + CMB + BAO, the best-fit values  are
$\Omega_{\rm{M0}}=0.271_{-0.014}^{+0.015}$, $n=
-0.22_{-3.27}^{+0.34}$, $\beta=0.74_{-0.56}^{+1.15}$, while with
GRBs + CMB + BAO, the best-fit values are
$\Omega_{\rm{M0}}=0.285_{-0.014}^{+0.015}$, $n=
-0.06_{-5.11}^{+0.59}$, $\beta=0.81_{-0.51}^{+3.80}$. We also
present the best-fit value of $\Omega_{\rm{M0}}$, $n$, and $\beta$
with 1-$\sigma$ uncertainties, and $\chi_{\rm min}^2$, $\chi_{\rm
min}^2/\rm{dof}$ for the  modified polytropic Cardassian model in
Table 1.

From Fig. 2 - 3  and Table 1,  we can find that GRBs can also give
strong constrains when combined to CMB and BAO data without SNe Ia.
By comparing to the joint constraints with GRBs and without GRBs, we
can see that the contribution of GRBs to the joint cosmological
constraints is a slight shift which adding the best-fit value of
$\Omega_{\rm M0}$, and significantly  narrowing  the parameters
confidence ranges of the modified polytropic Cardassian model.  We
also find that the $\Lambda$CDM model ($n\equiv0$, $\beta\equiv1$ )
is consistent with all the joint data in 1-$\sigma$ confidence
region, and combining these observational data can tighten the model
parameters significantly comparing to results from former works
(Wang 2007; Wang \& Wu 2010). We also investigate the deceleration
parameter for  Cardassian expansion models. The deceleration
parameter  $q(z)$ can be calculated by
\begin{eqnarray}\label {qH}
q=-1+(1+z)E(z)^{-1}\frac{dE(z)}{dz}.
\end{eqnarray}

In figure 4, we show the evolution of $q(z)$ for the original
Cardassian expansion model. We obtain $q_0=-0.55\pm0.054$, and the
transition redshift is $z_{\rm T}=0.73\pm{0.04}$ at the $1 \sigma$
confidence level, which is more stringent and comparable with the
former result ($z_{\rm T}=0.70\pm{0.05}$) by Wang (2007), but is
slightly later than the former result ($z_{\rm T}=0.55\pm{0.05}$) by
Wang \& Wu (2010) . We show the evolution of $q(z)$ for the
polytropic Cardassian expansion model in figure 5, and we find the
transition redshift $z_{\rm T}=0.68\pm{0.04}$ and
$q_0=-0.57\pm{0.07}$.

\begin{figure}
\begin{center}
\includegraphics[angle=0,scale=0.55]{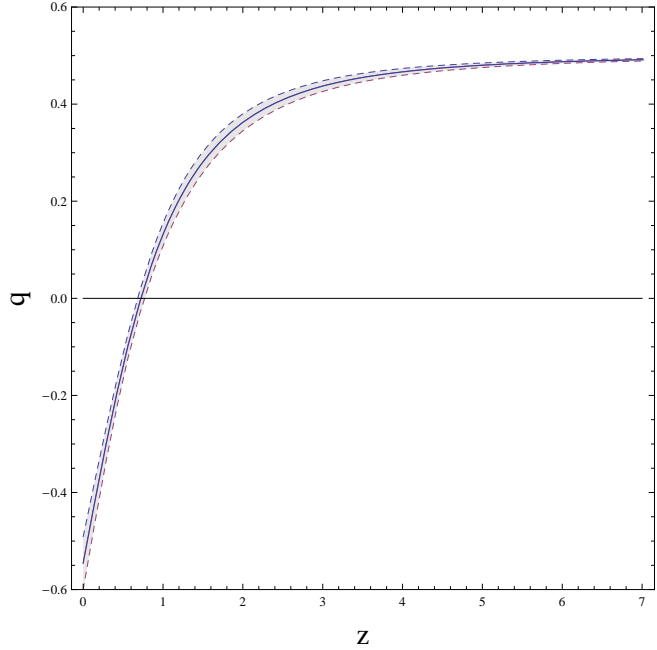}
\end{center}
\caption{The evolution of the deceleration parameter    $q(z)$ from
fitting results in the original Cardassian  model. The solid line is
drawn by using the best fit parameters. The shaded region shows the
1-$\sigma$ uncertainties.}
\end{figure}

\begin{figure}
\begin{center}
\includegraphics[angle=0,scale=0.55]{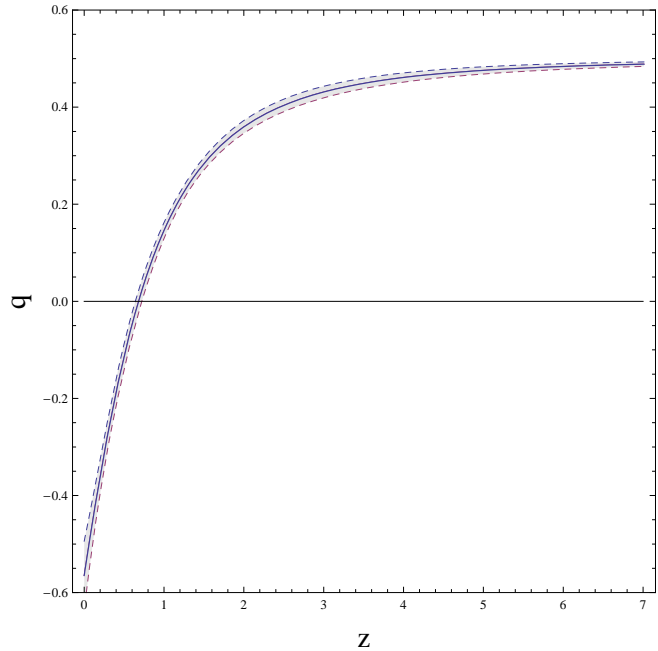}
\end{center}
\caption{The evolution of the deceleration parameter    $q(z)$ from
fitting results in the polytropic Cardassian  model. The solid line
is drawn by using the best fit parameters. The shaded region shows
the 1-$\sigma$ uncertainties.}
\end{figure}


\section{Conclusions and Discussions}
In this paper, by using the Union2 set of 557 SNe Ia, we calibrate
GRB data in a completely cosmology-independent way. When combine the
GRB data  with the Union2 set, we avoid any correlation between the
SNe Ia data and the GRB data (Liang et al. 2010). From GRB data to
the joint observations with the Union2 set, along with the CMB from
WMAP7 and the BAO observation from the SSDS Data Release galaxy
sample, we find significant constraints on  model parameters of the
original Cardassian model
$\Omega_{\rm{M0}}=0.282_{-0.014}^{+0.015}$, $n=
0.03_{-0.05}^{+0.05}$; and $n= -0.16_{-3.26}^{+0.25}$,
$\beta=0.76_{-0.58}^{+0.34}$  of the modified polytropic Cardassian
model, which are consistent with the $\Lambda$CDM model in
1-$\sigma$ confidence region. From the reconstruction of the
deceleration parameter $q(z)$ in Cardassian expansion models, we
obtain the transition redshift  $z_{\rm T}=0.73\pm{0.04}$ for the
original Cardassian model, and  $z_{\rm T}=0.68\pm{0.04}$ for the
modified polytropic Cardassian model, which are more stringent
comparing to the former results (Wang 2007;  Wang \& Wu 2010). It is
found that GRBs can  give strong constrains when combined to CMB and
BAO data without SNe data, and we can see that the contribution of
GRBs to the joint cosmological constraints by comparing to the joint
constraints with GRBs and without GRBs. Hereafter, along with more
and more observed data, GRBs could be used as an optional choice to
set tighter constraints on the Cardassian model and even other
cosmological models.

Recently, some works point out that there are observational
selection bias in GRB relations (Butler et al. 2007; Shahmoradi \&
Nemiro 2009) and possible evolution effects in GRB relations (Li
2007; Tsutsui et al. 2008). However, it is found that no sign of
evolution with redshift of the Amati relation, and the instrumental
selection effects do not dominate for GRB relations (Ghirlanda et
al. 2008, 2009). Nevertheless, further examinations of possible
evolution effects and selection bias should be required for
considering GRBs as standard candles for cosmological use.

\begin{acknowledgements}
This work was supported by the National Science Foundation of China
under the Distinguished Young Scholar Grant 10825313, the Key
Project Grants 10533010, and  by the Ministry of Science and
Technology national basic science Program (Project 973) under grant
No. 2007CB815401. PXW acknowledges partial supports by the National
Natural Science Foundation of China under Grant No. 10705055,  the
FANEDD under Grant No. 200922, the NCET under Grant No. 09-0144. NL
thanks Yun Chen, He Gao, Shuo Cao, Hao Wang, Yan Dai,  Fang Huang,
Jie Ma, Xinjiang Zhu, and Dr. Yi Zhang for discussions.
\end{acknowledgements}

\end{document}